# Phenomenology and physical origin of shear-localization and shear-banding in complex fluids

G. Ovarlez, S. Rodts, X. Chateau, P. Coussot

Université Paris-Est, Laboratoire Navier, 2 Allée Kepler, 77420 Champs sur Marne, France

**Abstract:** We review and compare the phenomenological aspects and physical origin of shear-localization and shear-banding in various material types, namely emulsions, suspensions, colloids, granular materials and micellar systems. It appears that shear-banding, which must be distinguished from the simple effect of coexisting static-flowing regions in yield stress fluids, occurs in the form of a progressive evolution of the local viscosity towards two significantly different values in two adjoining regions of the fluids in which the stress takes slightly different values. This suggests that from a global point of view shear-banding in these systems has a common physical origin: two physical phenomena (for example, in colloids, destructuration due to flow and restructuration due to aging) are in competition and, depending on the flow conditions, one of them becomes dominant and makes the system evolve in a specific direction.

## 1. Introduction

In recent years a lot of works evidenced shear-banding effects in complex fluids. This expression of shear-banding actually encovers a wide range of phenomena in which the shear rate profile in a flowing material exhibits an apparent discontinuity, i.e. at a given scale of observation the shear rate takes two significantly different values in two adjoining regions in which the shear stress is almost the same. This effect seems to occur with various types of systems ranging from fluids exhibiting a strongly non-Newtonian behavior (e.g. pasty fluids) to fluids with complex evolving structures (e.g. micellar solutions). There already exists a very consistent framework of knowledge concerning the physical origin and dynamical modelling of shear-banding in micellar systems. However the situation is quite different in the field of soft-jammed systems (colloids, emulsions, foams, gels) and granular pastes for which shear-banding has only recently emerged as an important issue. Several reviews on shear-banding were recently published, focused on the theoretical models [Fielding 2007, Dhont and Briels 2008, Olmsted 2008], the experimental techniques and measurement problems [Manneville 2008], and the specific bringing-in of NMR velocimetry and spectroscopy [Callaghan 2008]. In this paper our purpose is to review this field with the aim of (i) covering a wider range of materials including in particular soft-jammed systems and granular materials and (ii) distinguishing and clarifying the main phenomenological trends of this shear-banding and suggesting the main physical process at the origin of these trends. As a consequence from a theoretical aspect this paper is far behind modelling developments of shear-banding [Fielding 2007, Dhont and Briels 2008, Olmsted 2008] but we think that for various dispersed systems it is necessary to first clarify the trends in order to identify which of the existing theories is appropriate.

Let us recall that shear-banding was originally the word used to describe an effect observed with granular flows or soils [Nedderman 1992, Schofield and Wroth 1968]: either in simple shear or in a triaxial compression under an increasing force the soil or the granular mass is slowly deformed until a critical value beyond which it suddenly flows more rapidly with a shear localized in a thin region of thickness typically ranging from 5 to 10 grain sizes. Actually this is more generally a situation encountered with solid bodies [Cottrell 1964, Tabor 1991]: beyond a critical deformation a brittle solid breaks and a ductile solid deforms along specific surface as a result of dislocations. In both cases the deformation is localized in a thin region so that the apparent shear rate is discontinuous at the usual scale of observation. In this context it is not so surprising that shear-banding also occurs in pasty materials which are materials intermediate between solids and fluids.

In solid mechanics engineering science, shear banding is usually modelled as an instable response of the material leading to the apparition of discontinuity surfaces of the velocity or of the shear rate (Hill, 1952, Rice 1976). Rice (1976) has studied the conditions for the onset of shear-banding for many different constitutive laws, demonstrating that the existence of shear-banding cannot be related to a single feature of constitutive laws: they may present a minimum, a maximum, as well as a plateau. As an example, one can consider shear banding observed in thin sheets made up of ductile metals subjected to quasi-static uniaxial traction loading. In such a loading process, localization of deformation can appear just as well in the hardening regime (i.e. when the applied force is an increasing function of the prescribed displacement) as in the softening regime (i.e. when the applied force is a decreasing function of the prescribed displacement), depending on the material. Moreover, the condition for the onset of localization also depends on the boundary condition applied to the solid (prescribed displacements or prescribed forces) (Benallal et al 1989). It is also interesting to look at the situation of a thin sheet made up of a linear elastic ideal plastic material. When submitted to a increasing uniaxial elongation, the behaviour of this solid in a elongational strain-normal stress diagram is linear as long as the normal stress remains lower than the elastic yield stress and is described by a plateau when the normal stress reaches the elastic yield stress. While the solid deforms homogeneously in the elastic regime, it can be shown that non-trivial heterogeneous velocity fields can develop in the plastic regime, leading either to localized shear banding or to diffuse necked region. The nature of the localization of deformation actually depends on the shape of the elastic domain and the plastic flow rules, among others. Generally, it is observed that shear banding is favoured by elastic yield surface containing vertex (as Tresca or Mohr-Coulomb yield surface) and by non normal plastic flow rule. These features clearly indicate that conditions for the onset of shear banding relate to subtle features of the constitutive law of the material and that the tensorial nature of constitutive law describing continuous media's behaviour has to be taken into account to accurately model this phenomenon.

In fluids, shear-banding was first observed with micellar solutions then more recently with colloids (Fielding 2007). In parallel, in the continuity of the knowledge for granular materials the possibility of shear-banding was discussed for granular flows (Mueth et al. 2000) or foam flows (Kabla and Debrégeas 2003). It is now also a matter of discussion with concentrated suspensions of non-colloidal particles (Huang et al. 2005, Ovarlez et al. 2006) and emulsions (Becu et al. 2004). Actually this subject has taken advantage of the developments of new techniques for observing flow field inside transparent or non-transparent materials such as rheo-optical tools, scattering techniques and MRI (Magnetic Resonance Imaging) velocimetry (Manneville 2008, Callaghan 2008, Rodts et al. 2004).

However shear-banding may also be mistaken for another effect due to the strong non-linearity of the rheological behavior of some materials. More specifically this concerns yield stress fluids, which can exhibit a yielding (flowing) and an unyielded (static) regions in the same sample under various flow conditions such as Couette or capillary flows, leading to the well-known effect of "plug flow" in some regions where the stress is sufficiently small (Bird et al., 1982). At first sight this effect gives the appearance of two regions of very different shear rates, one in which the fluid is sheared and the other one where it is not sheared at all, which could be described as a kind of shear-banding. So here one of the main point of this paper is to clarify the differences and common features of these behavior types (i.e. shear-banding and solid/liquid coexistence due to yielding).

According to the above definition of shear-banding, if two regions flowing at different shear rates coexist in the same sample while the shear stress is not significantly heterogeneous the flow curve of the materials contains two parts associated with very different shear rates but with almost the same stress. This situation is represented in Figure 1 in the following way: in the region $[\tau_1;\tau_c]$ the shear rate is close to $\dot{\gamma}_1$; in the region $[\tau_c;\tau_2]$ the shear rate is close to $\dot{\gamma}_2$. In order to ensure the continuity of the curve there should be a region in which the stress decreases as the shear rate increases (see Figure 1). This situation is in fact unrealistic: a flow curve with a decreasing part cannot describe the rheological behavior of a single material in steady-state because the corresponding flows are necessarily unstable, as may be shown from a simple linear stability analysis [Tanner 1988, Coussot 2005]. Another possibility, which is in fact an asymptotic case of the previous situation, is that the flow curve takes the form of a plateau for $\tau = \tau_c$. In that case, in theory, various velocity fields can be obtained for the same imposed stress distribution. In practice one can expect that such a plateau would not be perfectly horizontal but anyway one may wonder whether this situation corresponds to a single state of matter and how the flow history can play a role on the possible evolutions of this state. This explains why it is so important to detect and understand shear-banding: (1) it may greatly influence the apparent flow properties (apparent flow at various velocities but almost constant stress) and (2) it can reflect time and spatial variations of the state of matter.

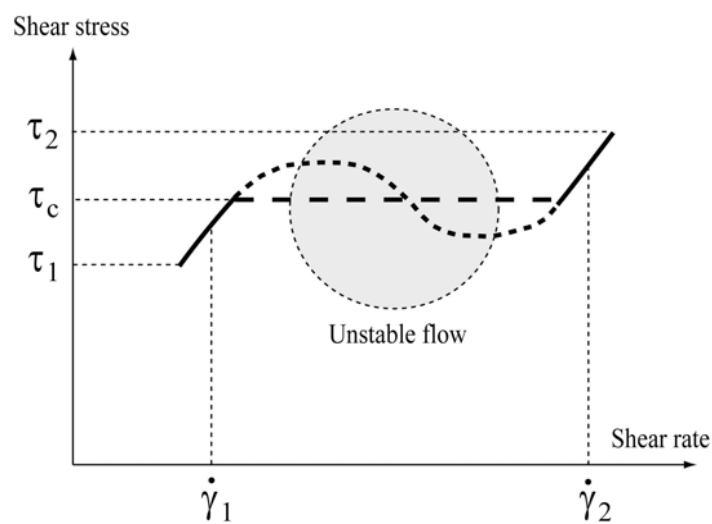

Figure 1: Need for a curve with a decreasing part (dotted line) (or a plateau (dashed line)) for connecting two parts associated with two different shear rates and a similar shear stress as a result of shear-banding.

Considering the critical role of stress in the definition of shear-banding we deduce that in order to analyze in depth shear-banding it is critical to control the stress distribution within the material. This is not the case for experiments with cone and plate geometry: the stress is generally almost uniform but we do not control the slight heterogeneities in the stress distribution, so that one can appreciate qualitatively the existence of shear-banding but one can hardly extract further information about the rheological behavior of the material. By contrast, with Couette flow, i.e. coaxial cylinders in relative rotation around their axis, the shear stress distribution is heterogeneous but well controlled. Under usual assumptions the shear stress is related to the torque $M$ applied to the inner cylinder and the distance $r$ from the axis: $\tau(r) = M/2\pi h r^2$ in which $h$ is the fluid height in contact with the inner cylinder.

In the following we start by considering the phenomenon of coexistence of static and flowing regions in yield stress fluids and examine its difference with shear-banding. Then we review the characteristics of shear-banding in various material types, i.e. pastes, granular pastes, micellar solutions, and in each case we review the observed trends, the basic modelling approaches and the probable physical origin. Note that concerning the modelling aspects we shall consider only the flow curve of the materials as it makes it possible to analyze and compare the different materials on a common basis but in some cases much more can be said on the dynamics modelling and the reader is referred to the reviews by Fielding (2007) and Olmsted (2008) for more details and references.

**2. Shear rate heterogeneity in yield stress fluids**

2.1 Observations

It is well-known that in many pasty materials flowing in a straight conduit one can observe an unsheared region far from the wall (the so-called plug) and a sheared region along the wall (see for example Figure 2). This is a typical trend observed with yield stress fluids [Bird et al. 1982]. At first sight this situation looks as a shear-banding: there is a flowing region and a rigid region and since the shear rate in the solid region is apparently equal to zero it can strongly differs from some average shear rate of the flowing region, which obviously differs from zero.

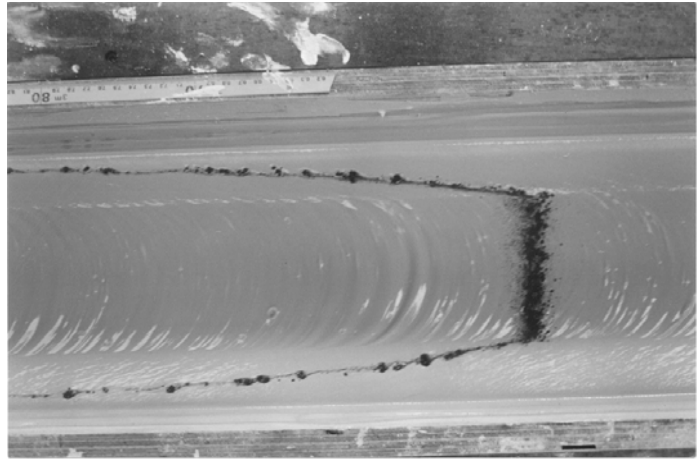

Figure 2: View from above of the free surface of a kaolin suspension flowing (here from left to right) in steady-state in an inclined, rectangular, open channel. A pepper line was dropped upstream perpendicularly to the flow direction. It now appears deformed due to the shear along the walls, and undeformed in the central plug region. Remark that the slight "dunes" at the free surface were formed far upstream at the hopper exit and do not play any role. [reproduced with Courtesy from APS, Coussot et al. 2002a]

However, conduit flows are also characterized by a strongly heterogeneous shear stress distribution. As a consequence, this separation between a liquid and a solid region may simply reflect the shear stress variation through the sample. In order to conclude for the presence or absence of shear banding, one has to look more precisely at the variation of the shear rate at a local scale in the sample, particularly at the approach of the interface between the liquid and the solid regions.

Such local measurements can be obtained thanks to MRI velocimetry in another geometry characterized by a strongly heterogeneous shear stress distribution: the Couette geometry. Two examples of velocity profiles inside yield stress fluids are shown in Figure 3 and 4. For this geometry the shear stress decreases by a factor 2.25 from the inner to the outer cylinder. In Fig. 3, we see that in a Carbopol gel, the shear rate, which is indirectly related to the slope of the tangential velocity profile via $\dot{\gamma} = r\,\partial(V/r)/\partial r$ in a Couette geometry, progressively decreases towards zero in the sheared region and then remains almost constant around zero in the apparently unsheared region. This defines respectively a liquid and a solid region, but there is no shear-banding as the shear rate progressively decreases from a finite value to zero and has the same value (zero) on both sides of the interface between the two regions: we have a coexistence of static and flowing regions.

Actually in order to have a strict appreciation of this continuous transition from the liquid to the solid region we would need to have a look at any scale, which obviously is not realistic, in particular because this would necessarily fail at the scale of the basic elements of the fluid. Thus, as usual in continuum mechanics we must appreciate the continuity of the shear rate at a scale sufficiently large for the continuum equation to be valid. This implies that the continuity of the shear rate likely fails at the approach of the scale of the material elements.

In the example of Figure 3 we show (see inset) that we get a similar aspect of the velocity when we decrease significantly the scale of observation, which confirms the continuity of the shear rate in our range of observation. Note that in this case anyway it was not realistic to go down to a smaller scale relevantly as this would fall below the resolution of the velocimetry technique.

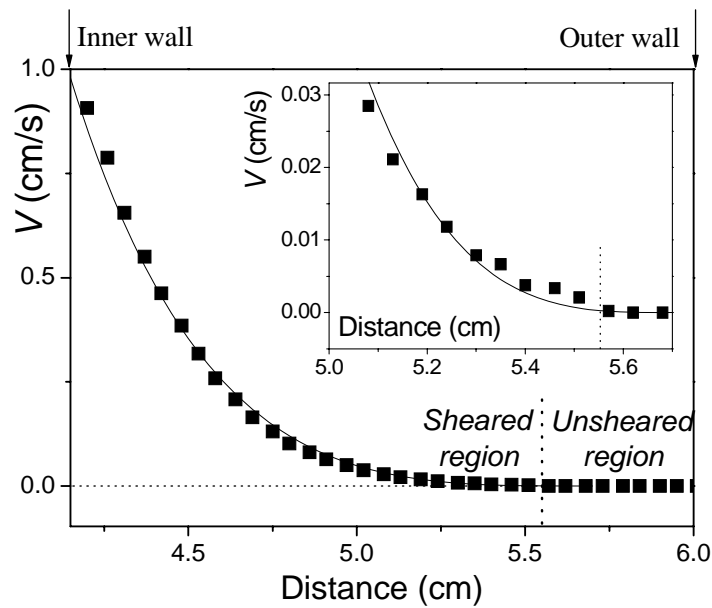

Figure 3: Steady-state velocity profile for Carbopol gel inside the gap of a Couette rheometer (inner diameter (including roughness): 8.3cm; outer diameter: 12cm) for a rotation velocity of the inner cylinder $\Omega = 3$ rpm (see experimental procedure in Coussot et al. (2009)). The inset shows the same data at a smaller scale, at the approach of the unsheared region. The continuous line (in the inset and the main figure) is a single Herschel-Bulkley model ($f = k\dot{\gamma}^n$ in equation (1)) fitted to the data with $n = 0.45$.

By contrast, we observe a discontinuity in the slope of the velocity profile measured during the flows of a cement paste in a Couette geometry in Figure 4: the velocity profile has an almost constant slope over a length significant with regards to the sample dimension, then, at a certain distance, the slope of the velocity vs distance curve abruptly drops to almost zero and keeps this value for larger distances. We still have the coexistence of a solid and a liquid region in such a flow. However here there is a discontinuity of the shear rate at the interface between the two regions: the shear rate is equal to a critical shear rate $\dot{\gamma}_c$ in the liquid region, and equal to zero in the solid region. This behavior is closer to what is usually described as shear-banding.

2.2 Modelling

The behavior of yield stress fluids in simple shear is generally modelled in the following way:

$$\tau < \tau_c \Rightarrow \dot{\gamma} = 0 \, ; \, \tau > \tau_c \Rightarrow \tau = f(\dot{\gamma}) \tag{1}$$

in which $\tau$ and $\dot{\gamma}$ are the shear stress and shear rate magnitudes, and $\tau_c$ the yield stress. We might also use a 3D description (see for example Bird et al. 1982) but for the sake of simplicity we will assume simple shear only, which in fact does not limit the generality of the purpose. In this formulation we have also left apart other aspects of the behavior of yield stress fluids such as the possible viscoelastic and thixotropic effects in the liquid and in the solid regimes. This is reasonable as we are dealing with steady-state flows.

In this context, coexisting static-flowing regions in geometries with heterogeneous shear stress distributions is simply the consequence of the existence of a yield stress. When the stress is larger than the yield stress within all the sample it is in its liquid regime everywhere and the shear rate continuously varies within the sample. By contrast, when somewhere the stress is smaller than the yield stress the material is in its solid regime (i.e. the deformation is limited) in this region and in its liquid regime in the rest of the sample. Finally the absence or the existence of shear banding associated with this coexistence of static and flowing regions only depends on the form of the flow function $f$. In the case of simple yield stress fluids $f^{-1}(\tau_c) = 0$, and $f$ is a continuous increasing function. Then the shear rate at the interface is zero on both sides of the interface. For complex yield stress fluids experiencing shear banding, there must be a discontinuity in the flow curve, i.e. $f^{-1}(\tau_c) = \dot{\gamma}_c > 0$. These two types of curves are represented on Figure 11.

2.3 Physical origin

The coexistence of static and flowing regions is associated with the existence of a yield stress. From a physical point of view a great variety of materials such as concentrated emulsions, foams, colloidal suspensions or physical polymeric gels can be considered as jammed systems [Liu and Nagel 1998] since they are made of a great number of elements (droplets, bubbles, particles, polymer chains) in strong (direct or at distance) interaction in a limited volume of liquid. From a mechanical point of view they can be considered as yield stress fluids, in the sense that they cannot flow in steady state unless the stress applied to them overcome a critical, finite value. This property results from the existence of a continuous network of interactions (i.e. jammed structure) between the elements, which has to be broken for flow to occur. The yield stress of the material is thus related to the strength of this network. This picture will be refined in Section 3 in order to explain why some yield stress fluids experience shear banding while others do not.

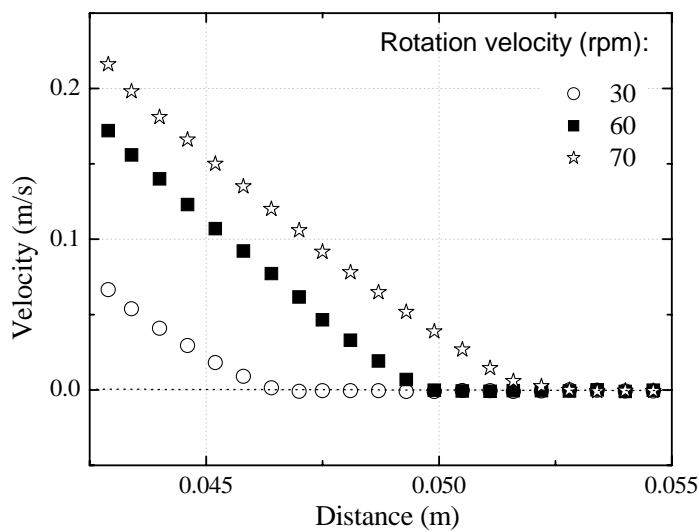

Figure 4: Velocity profiles under different rotation velocities in a Couette geometry (same as in Figure 3) for a cement paste (see experimental procedure in Jarny et al. 2005).

## 3. Shear-banding in soft-jammed materials

3.1 Observations

*Shear-banding in steady-state*

For various material types made of a high concentration of elements interacting at distance and thus exhibiting a yield stress (clay suspensions (Raynaud et al. 2002, Coussot et al. 2002a, Bonn et al. 2002), cement pastes (Jarny et al. 2005), emulsions (Bertola et al. 2003), foams (Lauridsen et al. 2004, Rodts et al. 2005)) we observe a discontinuity in the slope of the velocity profile in a Couette geometry. An example of this effect is shown in Figure 5: the velocity profile has an almost constant slope over a significant distance, then it drops to almost zero and remains around this value for larger distances. As a consequence the corresponding flow curve has the form of a curve truncated below a critical shear rate ($\dot{\gamma}_c$) associated with a critical shear stress ($\tau_c$): for a stress below $\tau_c$ no steady flow can occur, and for a stress larger than $\tau_c$ the shear rate is necessarily larger than $\dot{\gamma}_c$ (see example in Fig. 11).

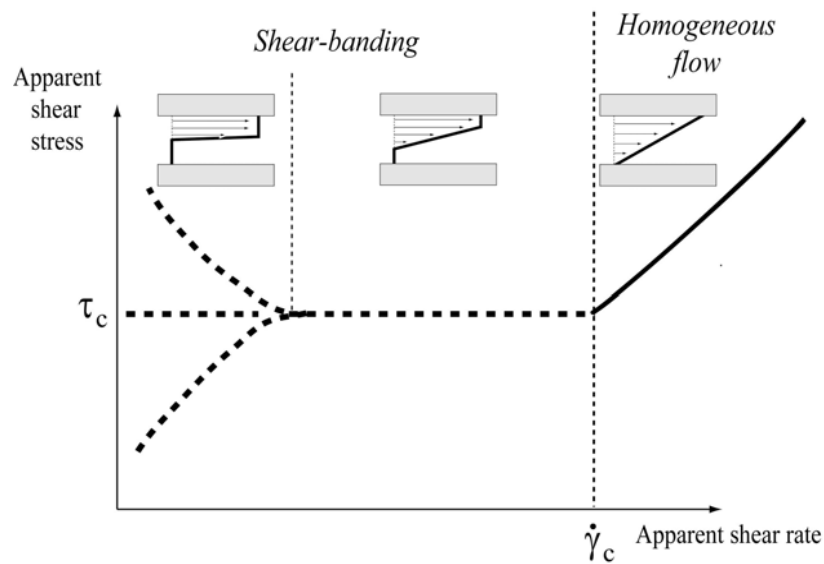

Figure 5: Typical flow curve obtained for some pasty materials under imposed, apparent shear rate: the solid line corresponds to the rheological behavior of the material in steady-state homogeneous flow; the dotted lines correspond to the various apparent steady-state flow curves that can be observed in practice for shear rate below the critical value ($\dot{\gamma}_c$), but which do not correspond to an effective bulk behavior of the material (see text). The drawings show the qualitative aspect of the velocity profile within the gap of the shear geometry in these apparent steady-state flows.

With such a material let us now consider the case of a slightly heterogeneous stress distribution centered around the mean value $\tau$. A steady flow is obtained only when in some region the stress is larger than $\tau_c$. In that case, since everywhere the stress is only slightly different from $\tau_c$ we have $\dot{\gamma} \approx \dot{\gamma}_c$ in some region and $\dot{\gamma} = 0$ in the rest of the material. This situation is typically that usually associated with shear-banding.

Let us now consider the situation for which an apparent shear rate $\dot{\gamma}$, smaller than $\dot{\gamma}_c$, is imposed to such a material in a geometry (with a gap $H$) in which the stress is almost homogeneous (typically a cone and plate geometry). The apparent stress will adjust so as to obtain a situation similar to that described in the previous paragraph: $\tau$ smaller than $\tau_c$ in some region, and larger than $\tau_c$ in the rest of the material; and the total thickness of the sheared regions is $h$ such that $\dot{\gamma}H = \dot{\gamma}_c h$, which makes it possible to get the imposed apparent shear rate. We get the picture presented in Figure 5. Such a result (with a cone and plate geometry) was obtained with different clay suspensions (Pignon et al. 1996, Coussot et al. 2002a) and with a silica gel (Moller et al. 2008).

A detailed illustration of this effect as it occurs with a bentonite suspension in a cone and plate geometry is shown in Figures 6 and 7: for a rotation velocity larger than a critical value the bands of shear for the different levels occupy the whole sample; for a smaller rotation velocity the bands of shear occupy only a limited thickness in the gap. Also the thickness of the sheared layer is constant for $\dot{\gamma} = \Omega/\alpha > \dot{\gamma}_c$ and decreases proportionally to $\dot{\gamma}$ for $\dot{\gamma} < \dot{\gamma}_c$ (see inset of Figure 7).

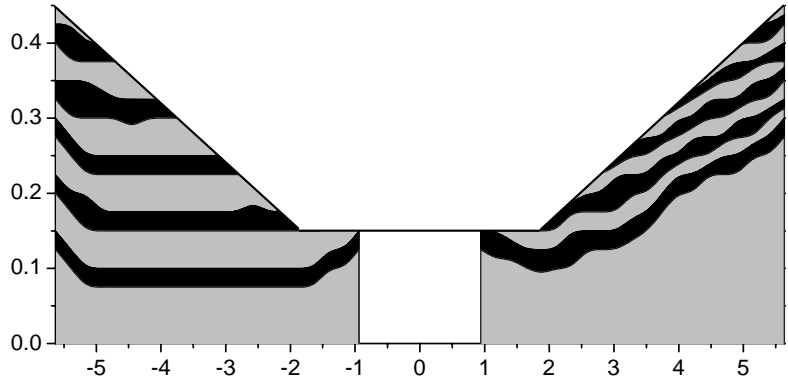

Figure 6: Velocity distribution in the gap of a cone and plate geometry (cone angle : 4.5°, diameter: 12cm) as observed by MRI velocimetry for a bentonite suspension (see technique, procedure and material characteristics in Raynaud et al. 2002). The vertical and horizontal scales are in centimetres. The material was initially presheared at a high velocity (say 110rpm) then the rotation velocity was fixed at a given level and the velocity distribution was measured during 5min after 10min of flow at this level. It was checked from other measurements at larger times that the steady-state had been reached. The left side shows the typical aspect of the velocity distribution (with a uniform shear) for a rotation velocity larger than 25rpm (here 80rpm) and the right side the typical aspect (with a localization of the shear in a band along the cone surface) for a rotation velocity smaller than 25rpm (here 10rpm). The successive bands from bottom to top represent the regions of velocities in 11 ranges of equal thickness and covering the complete range, i.e. $[0, V_M/10]$; $[V_M/10, 2V_M/10]$; etc, in which $V_M$ is the maximum velocity (along the cone surface).

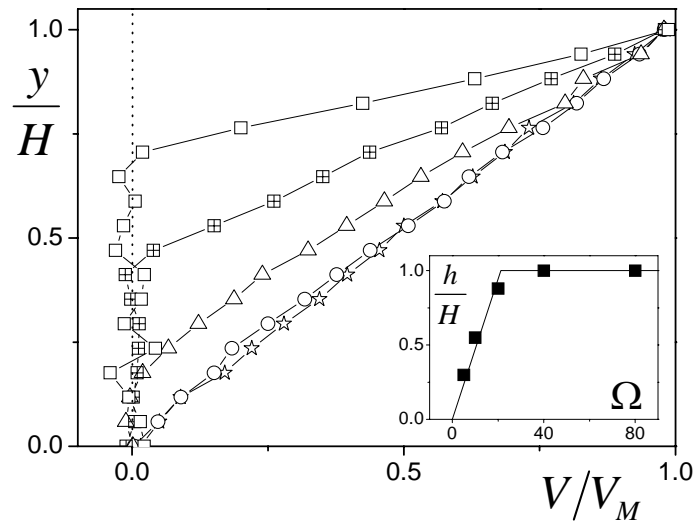

Figure 7: Local data for the experiments described in Figure 6 for different rotation velocities of the cone (stars: 80rpm; circles: 40rpm; triangles: 20rpm; crossed squares: 10rpm; squares: 5rpm): dimensionless height (with $H$ the gap thickness) above the plane as a function of the velocity scaled by the maximum velocity. The inset shows the sheared thickness (scaled by the gap) as a function of the rotation velocity.

*Discrete regime*

In general shear-banding does not have a significant effect from a macroscopic point of view as long as the sheared region is not too small: the apparent shear rate varies in a wide range while the apparent shear stress remains around $\tau_c$, a situation globally similar to that observed for a simple yield stress fluid. The situation is quite different when $\dot{\gamma} \to 0$ since in that case $h$ tends to zero and finally reaches a value of the same order as the size of the elements composing the material. In that case the continuum assumption is no longer valid, since the flowing region has a size not much smaller than the element size. It follows that the apparent behavior, i.e. the relationship between the apparent shear rate and the resulting shear stress, doesn't reflect the behavior of a single, homogeneous material, but more likely the behavior of a discrete material, i.e. which likely cannot be considered as a continuum material. This effect was observed with foams [Herzhaft et al. 2002, Rodts et al. 2005, Gilbreth et al. 2006], laponite suspensions [Pignon et al. 1996], emulsions [Ragouilliaux et al. 2006] and depending on the material structures either an increasing or a decreasing apparent flow curve were observed in the very low shear rate region (see Figure 5).

Note that more generally, a complex effect may be observed in all shear rate conditions if the flow is enforced in a small confined region. Actually, Goyon et al. (2008) have recently studied in detail this discrete regime, through the flows of emulsions confined in a microchannel. They observed that no local constitutive law can account for the flows of their system: the local behaviour of the emulsion seems to depend on the boundary conditions; they show that this may be the signature of a nonlocal constitutive law. This effect was observed in zones of extent that increase with the degree of jamming (the droplet volume fraction in this case) and that may involve up to 100 particles. Finally the physical origins and possible similarities with the "discrete" effects observed (see above) in the apparent flow properties still need to be clarified.

*Time-effects*

In fact the effects observed above do not develop instantaneously in the materials: time-effects play an important role. Let us consider a material exhibiting a critical shear rate as described above and in its liquid regime, i.e. the shear stress is everywhere larger than $\tau_c$. Then, at a given time we impose a shear stress smaller than $\tau_c$: the shear rate progressively decreases towards zero so that the material apparently stops flowing. In contrast, if the stress imposed is larger than $\tau_c$ a steady flow is obtained, but at a shear rate larger than $\dot{\gamma}_c$. This effect was described as a viscosity bifurcation [Coussot et al. 2002b] since instead of a progressive increase towards infinity as the stress is decreased towards $\tau_c$ the apparent viscosity reaches a finite value, i.e. $\tau_c/\dot{\gamma}_c$, just before the liquid/solid transition.

For such a fluid it is thus possible to obtain, at least transiently, a flow at a shear rate smaller than the critical value, simply by observing the flow as it is coming to a stop. This effect was observed under controlled shear stress by Ragouilliaux et al. 2007 via MRI velocimetry of an emulsion loaded with clay particles giving rise to attractive forces between the droplets. It can be appropriately represented in a rheogram through the time evolutions of the apparent (instantaneous) flow curve, i.e. $\tau(t)$ vs $\dot{\gamma}(t)$ (see Figure 8): (i) in a first stage the apparent flow curve a short time after preshear is more or less that of a power-law fluid (dashed plus continuous lines on the right of Figure 8), (ii) the points in the continuous part of the initial curve correspond to steady state flows, in this rheogram they remain almost fixed in time, (iii) the points situated in the dashed part of the initial curve do not correspond to steady-state flows, for a stress imposed in this range the shear rate progressively decreases towards very low values, so that after a sufficient time the apparent shear rate is zero, finally the flow curve in steady state in this range of stresses is the vertical, continuous line on the left in Figure 8.

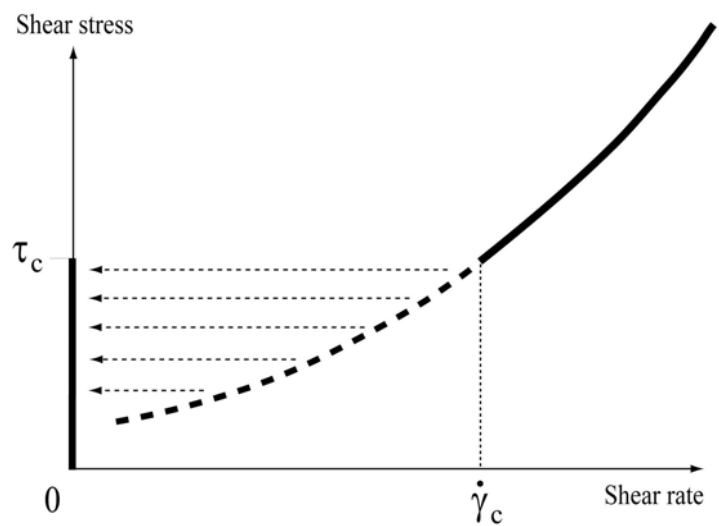

Figure 8: Typical flow curve obtained for some pasty materials (see text) under imposed, apparent shear stress: the continuous line corresponds to the effective (steady-state) flow curve; the dashed line is the apparent (transient) flow curve obtained after just a preshear at a high shear rate; the dotted arrows show the shear rate evolution from the apparent flow curve for different given shear stress values below the yield stress.

A slightly different evolution can be observed under controlled (apparent) shear rate (Rogers et al. 2008). In that case, after a strong preshear the initial apparent flow curve corresponds to that observed initially under controlled shear stress (cf Figure 8). Then for apparent shear rate larger than $\dot{\gamma}_c$ the stress remains almost fixed, this corresponds to the continuous line in Figure 8, i.e. homogeneous steady-state flows; for apparent shear rate smaller than $\dot{\gamma}_c$ the stress more or less rapidly increases towards the yield stress plateau (or to another value if the sheared thickness in the shear-band is too small (cf. § Discrete regime)) as shown in Figure 5. We thus get an apparent flow curve moving vertically from its initial value (dashed line of Figure 8) to its final value as shown in the case of granular pastes in Figure 9. However we emphasize that in that case, i.e. under controlled (apparent) shear rate, we get an apparent steady-state flow curve for which steady state flows below $\dot{\gamma}_c$ actually corresponds to heterogeneous flows (with shear-bands) and the apparent shear rate does not correspond to the local shear rate.

3.2 Modelling

In comparison with micellar systems (see below) modelling the rheological behavior of pastes with the aim of representing the above effects has been the subject of relatively few published works. One reason is that the phenomenon has only recently emerged. Another reason is that the structures of the systems exhibit a great variety, so that it is difficult to suggest a model at the same time sufficiently general and accurate.

For modelling steady-state it seems that the most natural way is to use a model similar to equation (1) but now with a function $f$ such that $f^{-1}(\tau_c) = \dot{\gamma}_c > 0$. Such a model effectively implies that no steady flows can be obtained below the critical shear rate $\dot{\gamma}_c$. In order to reflect the time-effects described above it is necessary to introduce a structure parameter which will represent some physical characteristics of the instantaneous structure state. In this picture, the constitutive law depends on this structure parameter, and the structure parameter evolution is driven by some kinetic equation. Then various thixotropy models can be suggested which predict the flow instability below $\dot{\gamma}_c$ [Coussot et al. 1993, Cheng 2003, Roussel et al. 2004]. These models predict that when the material has been presheared its structure parameter is low, so that an apparent flow curve as shown in Figure 8 can be obtained. Then under controlled stress above the yield stress the structure parameter reaches a steady-state value and a steady-state flow is obtained, whereas below the yield stress the structure parameter increases, so that the shear rate decreases, and eventually tends towards a large value for which no flow occurs.

3.3 Physical origin

The different trends above finally suggest a physical explanation for the viscosity bifurcation effect: usually the thixotropic character of these fluids is represented via a structure parameter which evolves as a result of a competition between restructuration effects (which are for example clearly observed at rest, the apparent viscosity increases in time) and destructuration effects due to the flow; it is reasonable to consider that for a sufficiently low stress this competition is won by the restructuration effects, which tend to decrease the shear rate, decreasing further the destructuration effects; on the contrary, for larger stress the competition is won by destructuration effects, so that the shear rate can increase, further destructuring the material until a steady-state is reached. Such a scheme should be general for jammed materials, and thus would be expected for any yield stress fluid since these materials have some structure in their solid regime which requires some characteristic time to form or break. However, as described in the previous section, there seems to exist some fluids for which these effects are not observable under usual flow conditions, i.e. the critical shear rate, if it exists, is very low. Finally it was recently suggested (Coussot et al. 2009) that two types of jammed materials can be distinguished: systems with mostly repulsive interactions between the elements and systems with mostly attractive interactions between the elements.

The analysis is as follows. In Carbopol gels (Coussot et al. 2009), repulsive latex suspensions (Wassenius and Callaghan 2005) and non-adhesive emulsions (Becu et al. 2006, Ovarlez et al. (2008)), the elements interact via different types of forces: colloidal forces between latex particles, forces due to surface energy storage between emulsion droplets, and elastic forces due to blob deformation in the gel. All these forces give rise to repulsive interactions when two neighbouring elements tend to get closer to each other, so that beyond a certain concentration of elements in the liquid they are jammed against each other: due to the interactions with its neighbours each element is in a potential well from which it can be extracted only by applying a force larger than a finite value. This structure starts to flow when somewhere in the material one element gets out of its potential well. After such an event the whole structure rapidly rearranges under the action of the local, elastic, repulsive forces. Remark that this picture corresponds to that suggested within the frame of the SGR (Sollich et al. 1997) or the STZ (Falk and Langer 1998) models. Thus a macroscopic deformation involves a succession of such local events followed by an almost instantaneous global rearrangement, which should, on average, leads to a homogeneous deformation. Finally the material apparently does not give rise to any shear-banding beyond the yield stress at least in our usual range of observation of shear rates.

Let us now consider systems with significant attractions between elements (colloidal suspensions, foams (since the bubbles are somewhat stuck to each other via very thin liquid film)). When they are dispersed at random in the liquid the elements develop some interactions with their neighbours so that, similarly to the case of materials with repulsive interactions, they can be considered as in a potential well due to their interactions with the surrounding elements. However, here the spatial distribution may significantly evolve in time under the combined action of thermal agitation and attractive forces. The elements may eventually find an arrangement in which they are linked to the others by attractive forces. In this new arrangement the potential well is in general significantly deeper than in the initial disordered dispersion. A macroscopic flow implies that some links are broken somewhere in this structure, but now the corresponding elements are for some time in a shallow potential well from which they can more easily escape than the elements still linked with their neighbours. Thus the broken links form weaker regions which can be more easily deformed subsequently and a shear-banding develops in these regions of lowest viscosity.

At last note that another trend was observed with more or less similar systems, i.e. crowded colloidal star polymer, which also exhibit an apparent yield stress: time fluctuations in the velocity profile varying with the flow history. It was suggested that these fluctuations are due to intermittent changes due to jamming/unjamming transitions (Holmes et al. 2004). This effect, which is reminiscent of those observed within shear-banding micellar systems (see Section 5), suggests that there might be some analogy in the physical origin of shear-banding in micelles and pasty materials.

## 4. Granular pastes

4.1 Observations

Interestingly similar effects apparently occur with a system in which the particles in suspension do not develop any interaction others than hydrodynamic forces and contact forces. From a macroscopic point of view, these dense suspensions of noncolloidal particles seem to behave as simple yield stress fluids (cf. steady state flow curve in Figure 9). However, experiments performed in a controlled stress mode reveal that they undergo a viscosity bifurcation (Huang et al. 2005, Ovarlez et al. 2006) below the apparent yield stress: no flow seems to be allowed below a critical shear rate. This is confirmed by local measurements of the velocity profiles in a Couette geometry: when the apparent shear rate (or rotational velocity) falls below the critical shear rate evidenced in the viscosity bifurcation experiments, the flow is localized near the inner cylinder to ensure that the shear rate is higher than the critical shear rate in the flowing region (Huang et al. 2005, Ovarlez et al. 2006).

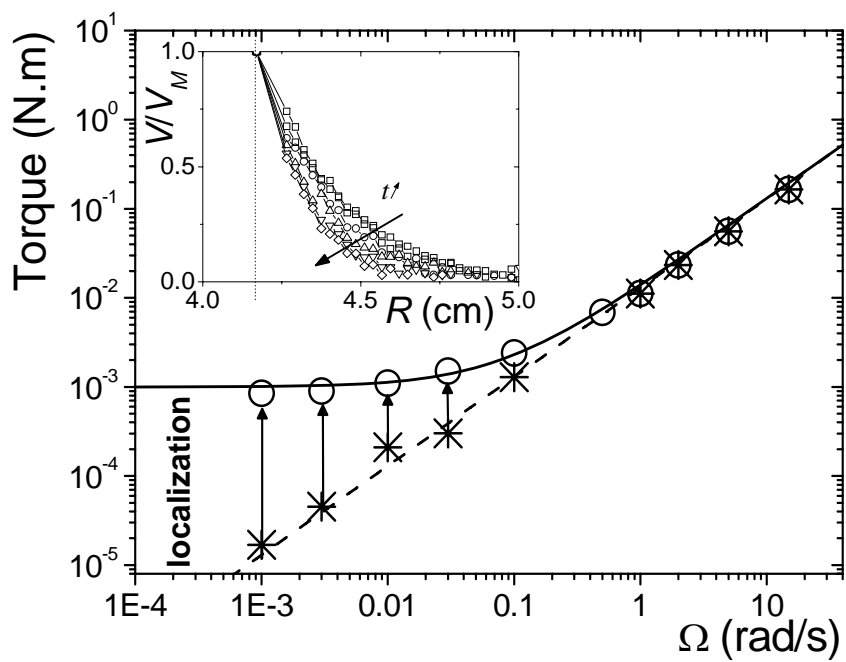

Figure 9: Torque vs. rotational velocity for the flows of a non-colloidal suspension in a Couette geometry (see material and procedure in Ovarlez et al. 2006). Open circles: steady state. Stars: torque measured instantaneously after a velocity step from a velocity of 100 rpm. The inset shows the velocity profile in time when the rotation velocity is changed from 9rpm to 0.2rpm at $t = 0$. The profiles shown correspond to $t = 2.3s; 6.9s; 16.1s; 29.9s; 66.7s$.

The way shear localization develops is reminiscent of the observations made in pasty materials: from MRI it was shown that when the velocity of the inner cylinder is abruptly decreased below its critical value, the flow progressively stops in a region near the outer cylinder (Ovarlez et al. 2006) (see Figure 9). In parallel, the torque increases up to its plateau value corresponding to the yield stress. Thus, here again no flow is allowed below a critical shear rate and the constitutive equation in steady state should be written as a "truncated" viscous law. It has to be noted that the torque value at the onset of localization is proportional to the inner cylinder rotational velocity whatever this velocity (Ovarlez et al. 2006): this suggests that the behaviour of the flowing material is that of a purely viscous material, without any signature of a yield stress. Note that this contrasts with the observations with some pasty materials (see above) in which the apparent behavior just after preshear seems to be already that of a yield stress fluid [Ragouilliaux et al. 2007].

4.2 Physical origin

As the only interactions between particles may be hydrodynamic interactions and contact forces, we can give the following sketch: in the flowing region, the hydrodynamic interactions dominate as the behaviour is basically purely viscous; on the other hand, there must be a contact network (with frictional contacts between the particles) in the jammed zone as this zone must be able to sustain a minimum shear stress without flowing. The shear banding instability observed in these materials is thus likely the signature of a change in the nature of the interactions between particles. It has to be noted however that, thanks to the MRI display, concentration measurements were performed during the flows of these materials, and that no macroscopic sedimentation was observed during shear localization (Ovarlez et al. 2006). This would mean that the transition occurs at the contact scale between lubricated close contacts and frictional contacts. The shear banding instability would then occur when the force due to lubrication is lower than the force due to gravity, leading to the formation of contacts; this is reminiscent of the shear resuspension mechanism (Acrivos et al. 1993). This would require imperfect density matching between the particles and the fluid: this was confirmed by (Fall 2009). This mechanism is also reminiscent of the behavior of pasty materials in which shear banding is associated with the creation/destruction of adhesive contacts: here it is associated with the creation/destruction of frictional contacts.

At very low shear rates, the flow is strongly localized and the flowing region is a few particles wide. Then the behaviour resembles that of a concentrated granular paste (Ancey and Coussot

1999), and might be considered as a "discrete regime" as described in section 3.1. It is also reminiscent of the behavior of dry granular materials or foams which in Couette systems (Mueth et al. 2000, Da Cruz 2004, Gilbreth et al. 2006) exhibit velocity profiles localized along the inner cylinder while some residual motions are observed farther. In that case an original trend is the similarity of the profiles when the velocity is scaled by the maximum velocity (along the inner cylinder). It was shown recently (Rodts et al. 2005) that when this scaling occurs the material behavior in simple shear should follow a power-law model, which is not consistent with the apparent behavior of such granular materials at a macroscopic scale. Thus that case has a strong analogy with the discrete regime of pasty materials since no consistent constitutive equation can represent the apparent behavior.

Note that another source of shear localization exists in non-colloidal suspensions. When the volume fraction is too high, it was observed that the material separates into two zones, whatever the rotational velocity: a jammed zone of concentration higher than a maximum packing fraction (equal to 60.5% in monodisperse suspensions of spherical particles) near the outer cylinder, and a flowing zone of concentration lower than this maximum packing fraction near the inner cylinder (Ovarlez et al. 2006). In this case the structural change associated with shear banding is thus a change in the volume fraction. No critical shear rate can be defined in this case: the shear rate at the interface between the flowing zone and the jammed zone can take any value (which should still be above the critical shear rate evidenced above for the formation of frictional contacts) as the jammed zone can apparently never be enforced to flow.

**5. Micellar systems**

5.1 Observations

The flow curve of various wormlike micelle or lamellar phase solutions (Rehage and Hoffmann 1991, Roux et al. 1993, Cappelaere et al. 1997, Hernandez-Acosta et al. 1999, Holmqvist et al. 2002) was shown to exhibit a stress plateau in the shear stress vs shear rate curve at a particular stress value, i.e. the stress appears to remain almost constant in a certain range of shear rates. In fact the stress may slightly increase in this region but this increase appears negligible in comparison with its variations at shear rates out of this range (see Figure 9). Under controlled stress experiments, when progressively increasing the stress level, one observes around a critical value ($\tau_c$) a large increase of the resulting shear rate which rapidly turns from a small value ($\dot{\gamma}_1$) to a much larger value ($\dot{\gamma}_2$) associated to the end of the plateau. If the slow flow at shear rates smaller than $\dot{\gamma}_1$ has not been detected the fluid may seem to start to flow abruptly at $\tau_c$. In several cases it has also been shown that the position of the plateau is not precisely defined: in a portion of flow curve before the plateau and a portion after the plateau the flows are extremely stable, but at the approach of the plateau different shear stress vs shear rate data may be obtained depending on flow history (Bonn et al. 1998, Salmon et al. 2002) (see Figure 9). This in particular implies that flow curves obtained under an increasing or a decreasing ramp of stress may differ around the plateau. It was suggested that the flow curve of such material has a S-shape (as shown in Figure 1). Let us recall that for a S-shape curve a flow can be obtained either in the first or the second increasing part of the flow curve, but no stable flow can be obtained in the decreasing part.

Various techniques (Small Angle Neutron Scattering, Small Angle X-ray Scattering, birefringence,..) have been used to study the suggested structural transition in wormlike micelles exhibiting this peculiar rheological behavior. The observations in general led to the conclusion that the plateau is associated to a phase transition, such as the transition between isotropic and nematic phases in wormlike micelles (Schmitt et al. 1994, Berret et al. 1994), the coexistence of the lamellar phase and onions in other surfactant mixtures (Partal et al. 2001) or the coexistence of different orientational structures (Holmqvist et al. 2002). Since such measurements concerned the whole material or at least all the material in a volume across the gap, they provided a global information and it could be conceived that the phase transition was progressive in time but approximately homogeneous in the bulk at any time. However, from observations by flow birefringence on wormlike micelles it was finally noticed that this transition occurred in space (Cappelaere et al. 1997): in a Couette system the second phase appears at the first critical shear rate ($\dot{\gamma}_1$) and progressively invades the gap as the apparent shear rate ($\dot{\gamma}$) increases, until completely occupying the gap for $\dot{\gamma} = \dot{\gamma}_2$. With similar techniques and analogous materials it was confirmed that the shear rates in the two phases strongly differ [Cappelaere et al. 1995, Decruppe et al. 1995, Makhloufi et al. 1995, Berret et al. 1997].

These observations have encouraged people to focus on the velocity field within flows of soft jammed systems with the help of different techniques (NMR, light scattering, simple microscopy). All the results show that "shear-banding" develops, in particular in the plateau region (Mair and Callaghan 1997, Britton and Callaghan 1997, Salmon et al. 2003, Holmes et al. 2003).

5.2 Modelling

Assuming that we can leave apart time effects a simple means to represent these observations is to consider that the fluid in steady state flow can either be in a state (i) with a behavior in simple shear of the form: $\tau < \tau_c \Rightarrow \tau = f(\dot{\gamma})$; or in a state (ii) with a behavior of the form: $\tau > \tau_c \Rightarrow \tau = g(\dot{\gamma})$; in which $f$ and $g$ are two increasing, positive functions such that $f^{-1}(\tau_c) = \dot{\gamma}_1 < \dot{\gamma}_2 = g^{-1}(\tau_c)$. The apparent flow curve of such a fluid has a plateau between $\dot{\gamma}_1$ and $\dot{\gamma}_2$ like that described in Figure 10. Indeed, let us consider a situation for which some shear rate is imposed to the fluid while the shear is approximately homogeneous (for example in a cone and plate geometry). For an apparent shear rate $\dot{\gamma}$ in the range $[\dot{\gamma}_1; \dot{\gamma}_2]$, the stress is necessarily equal to $\tau_c$ otherwise the shear rate would be given by one of the above constitutive equations and would be outside the range $[\dot{\gamma}_1; \dot{\gamma}_2]$. The shear rates of the material in state (i) and (ii) are respectivey $\dot{\gamma}_1$ and $\dot{\gamma}_2$ and the sample develops two parallel bands in the state (i) and (ii) of respective thicknesses $h_1$ and $h_2$. The relative velocity of the boundaries may then be written $V = h_1\dot{\gamma}_1 + h_2\dot{\gamma}_2 = H\dot{\gamma}$, from which we deduce that, in the absence of wall slip or other perturbating effects, the corresponding relative gap fractions of the material in state (i) and (ii), respectively $\varepsilon_1 = h_1/H$ and $\varepsilon_2 = h_2/H$, should respect the relation:

$$\dot{\gamma} = \varepsilon_1\dot{\gamma}_1 + \varepsilon_2\dot{\gamma}_2 \tag{2}$$

This is the so-called "lever rule".

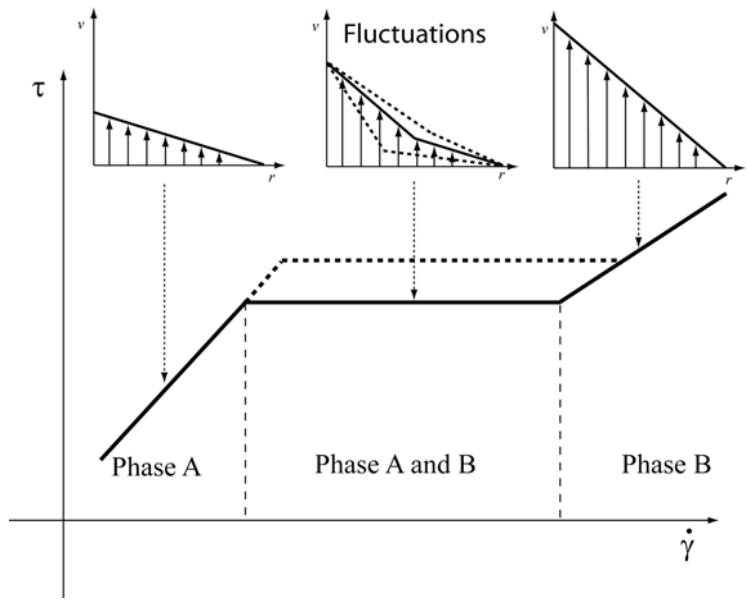

Figure 10: Schematic aspect of the velocity profiles for micellar solutions in Couette flows under different rotation velocities of the inner cylinder as observed by Salmon et al. (2003) and corresponding state of the material (in two possible phases). The continuous lines correspond to the velocity profiles averaged over some time and the dotted lines to the "instantaneous" velocity profiles. Below corresponding flow curve with the plateau phenomenon: the position of the plateau and the values of the critical shear rate may depend on flow history, leading to apparent flow curves situated between the continuous and the dashed lines.

5.3 Physical origin

The above results have been obtained under the assumption that the stress is perfectly homogeneous within the gap. In reality there always remain some sources of heterogeneity, which will determine the spatial distribution of the regions (i) and (ii). The state (i) thus preferentially localizes in the regions of smaller stresses and the state (ii) in the other regions. For micellar solutions, in contrast with pasty materials, the two coexisting regions appear to contain two materials of very different mesoscopic structures (Salmon et al. 2003) and various recent studies have suggested that the phenomena were more complex than the simple, stable shear-banding as described above. For example it has been shown that the band stability is questionable, the localization of shear in the regions of larger stresses does not seem obvious for all systems (Fischer and Callaghan 2000), and the "lever rule" implies an extreme localization of shear at low velocities, which is not in agreement with birefringence observations. Generally speaking the shear-banding has been shown to be a dynamical process (Decruppe et al. 2001), with an evolution depending on the flow history. For wormlike micelles it was also observed (Lerouge et al. 2000) from the study of the spatial distribution of the transmitted light intensity through the gap that in one band the flow can be inhomogeneous at a small scale (say 150 $\mu$m): the band is made of small sub-bands closely aligned in the flow direction. Finally the most complete studies (Salmon et al. 2003, Holmes et al. 2003) on this subject showed that the shear-banding is stable only when averaging the velocity profiles over sufficiently long times, but instantaneous velocity profiles (taken over about one second) appear to significantly fluctuate around this mean value. In particular the position of the interface between the two regions significantly varies in time (see Figure 10).

## 6. Conclusion

Here we suggested a clarification of the difference between shear-banding, for which there is an apparent discontinuity in the shear rate vs shear stress curve and the coexistence of static-flowing regions for which there is a continuous variation from two regions of different mean shear rates with a zero shear rate in one of them (see Figure 11). In this context, the perfect plastic behavior is situated at the interface (see Figure 11) between these two behavior types: for a critical stress the apparent shear rate can take any value so that it is not possible to decide whether or not we have a discontinuity of the shear rate.

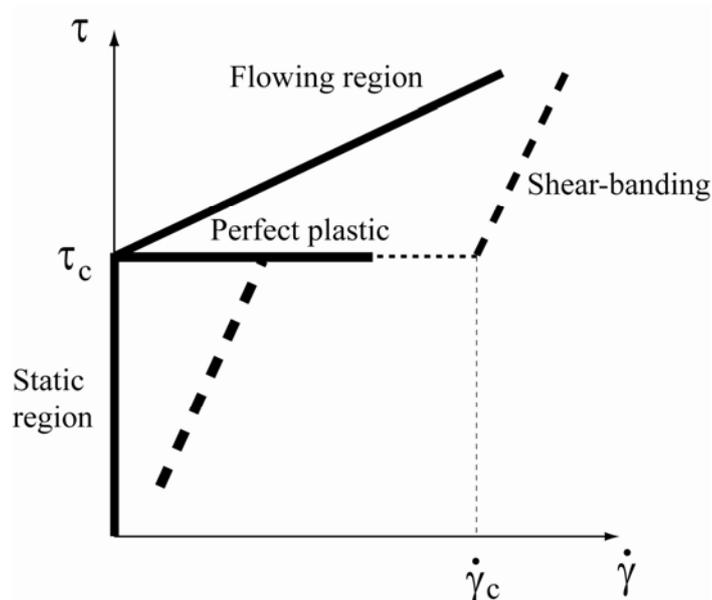

Figure 11: Main qualitative aspects of the shear rate vs shear stress curve for materials giving rise to shear-localization or shear-banding. The continuous line corresponds to a simple yield stress fluid in which a static region coexists with a flowing region. The dashed line corresponds to shear-banding. The vertical line corresponds to a perfect plastic behavior, intermediate between both behavior types.

Apparently various physical phenomena are at the origin of shear-banding in the different materials we reviewed. Actually if we look at them from a very global point of view there appears some analogies: the spontaneous restructuration within a colloidal system, which can overcome the destructuration due to shear for sufficiently low stresses; the formation of contacts due to sedimentation in granular pastes which is counterbalanced by hydrodynamic resuspension, but may ultimately lead to a jamming of the structure at low shear rates; the phase changes in micellar systems leading to regions of very different viscosities. In all these systems there are at least two physical phenomena which are in competition and depending on the flow conditions one of them becomes dominant and make the system evolve in a specific direction. In some sense this is consistent with the general description of Olmsted (2008): "When an imposed shear rate exceeds a characteristic structural relaxation time, the fluid can attain a nonequilibrium state whose structure is qualitatively different from that of the quiescent state".

However there exists some distinction between the physical trends occurring with micellar systems and with jammed systems (soft-jammed systems, granular pastes). Indeed for micellar systems the underlying phenomenon is a priori a kind of physical phase change, which occurs more or less immediately under given, local conditions. In contrast, with appropriate flow history jammed systems are able to flow homogeneously at any shear rate, at least over short duration, without developing shear-banding. These are the specific boundary conditions which lead them to develop shear-banding in time, as a result of the progressive temporal evolution of some structural characteristics of the material.

Finally shear-banding occurs in a wide variety of materials which exhibit a strongly non-linear rheological behavior or/and a complex structure. This suggests that for such materials shear-banding is not an exotic phenomenon but an almost general natural consequence, and that for complex, strongly non-linear materials this is the absence of shear-banding which is unexpected.